\documentclass[conference]{IEEEtran}
\IEEEoverridecommandlockouts
\usepackage{cite}
\usepackage{amsmath,amssymb,amsfonts}
\usepackage{algorithmic}
\usepackage{graphicx}
\usepackage{textcomp}
\usepackage{xcolor}
\usepackage{comment}
\usepackage[nottoc]{tocbibind}

\def\BibTeX{{\rm B\kern-.05em{\sc i\kern-.025em b}\kern-.08em
    T\kern-.1667em\lower.7ex\hbox{E}\kern-.125emX}}
\begin{document}

\title{Music Genre Classification: Training an AI model

}

\author{\IEEEauthorblockN{1\textsuperscript{st} Keoikantse Mogonediwa}
\IEEEauthorblockA{\textit{Academy of computer science} \\
\textit{University of Johannesburg}\\
Auckland Park JHB 2092, South Africa \\
}

}

\maketitle

\begin{abstract}
Music genre classification is an area that utilizes machine learning models and techniques for the processing of audio signals, in which applications range from content recommendation systems to music recommendation systems. In this research I explore various machine learning algorithms for the purpose of music genre classification, using features extracted from audio signals.The systems are namely, a Multilayer Perceptron (built from scratch), a k-Nearest Neighbours (also built from scratch), a Convolutional Neural Network and lastly a Random Forest wide model. In order to process the audio signals, feature extraction methods such as Short-Time Fourier Transform, and the extraction of Mel Cepstral Coefficients (MFCCs), is performed. Through this extensive research, I aim to asses the robustness of machine learning models for genre classification, and to compare their results.
\end{abstract}


\section{Introduction}

Music is a form of expression, a universal language that is easy to translate into cultural stories and different emotions. Communities and societies all over the world unite through music. The emergence of digital music in the industry has resulted in music being easily accessible from any technology device \cite{stafford2010music}. Online streaming platforms consist of different types of musical genre's. This research paper studies a gap in this industry, where genre classification algorithms are developed to categorize musical genres as a added feature to enhance a user's experience. Genre classification can be used in different areas within the music industry, such as assisting music producers in identifying which musical genres are more popular. 

A significant amount of work has gone into music genre classification models. Genre classification models that have been developed depend on the quality of the provided labels as part of the training process. Even though these models are advanced, there remains a gap in real world applications, where some data may not contain the necessary text labels required for genre classification \cite{buisson2022ambiguity}.

\subsection{Is the problem solved?}
Classification models mainly follow a supervised learning approach, thus meaning they require text labels that are clear and easily definable. The study of this paper, aims to cover an area where ambiguity could exist within genre classification. The niche research area is to establish if different machine learning algorithms can perform more accurately if they use input data extracted as features from audio files, instead of using text labels.

\subsection{A possible solution}
A possible solution is to use "Short-Time Fourier Transform" (STFT). STFT is the practice of using audio signals to process and analyse the frequencies of an audio file over a certain time period \cite{portnoff1980time}. Fourier transformation consists of different concepts such as overlapping windows, which is a practice that helps to capture smoother transitions between the time segments. This study proposes the use of Short-Time Fourier Transformation to extract features from an audio file. The features are then used as an input into different machine learning algorithms that are trained to classify audio files into their respective genre. 

\begin{figure}
    \centering
    \includegraphics[width=.5\textwidth]{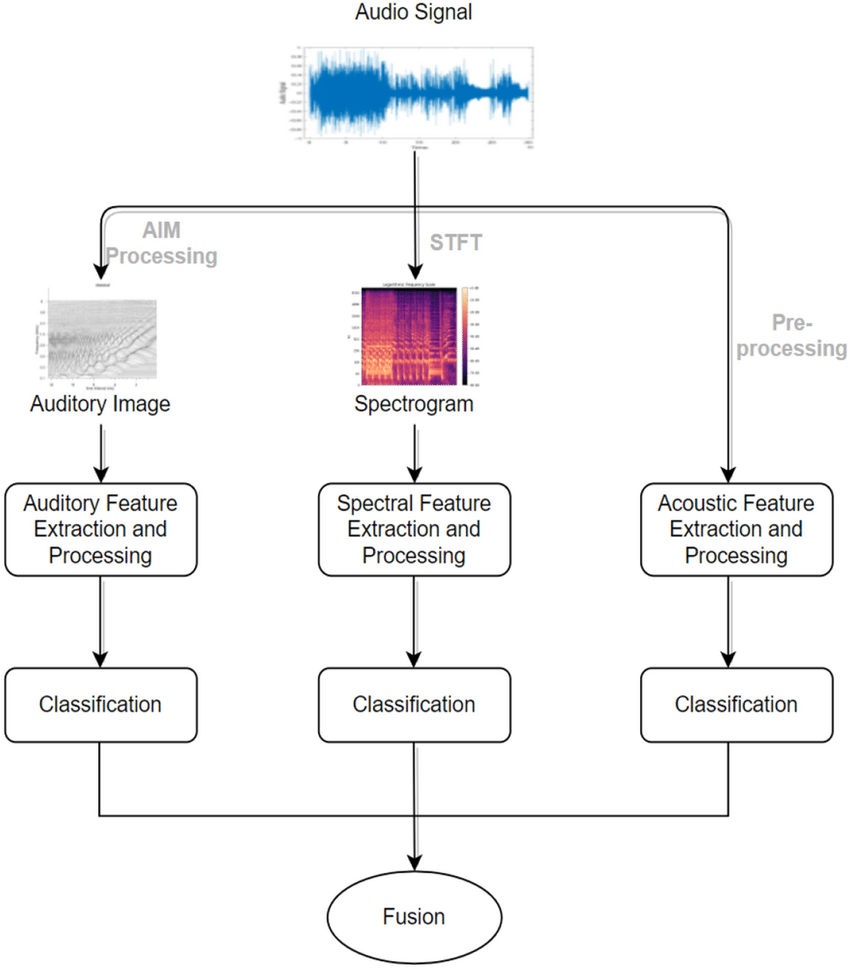}
    \caption{Typical flowchart demonstrating the steps involved in music genre classification}
    \label{fig:enter-label}
\end{figure}

\section{Intended Experiment methods}
Music genre classification within the machine learning niche creates a specialized research area that aims to develop algorithms that can automatically detect and categorize music. The algorithms include the Multilayer Perceptron, a Convolutional Neural Network (CNN), which is part of the deep learning models that are applied for various tasks such as computer vision or classification, the K-Nearest Neighbour (K-NN), and finally a Random Forest wide model that incorporates feature engineering. 

The four models are compared to identify which model produces the most accurate results. This research paper provides a deep understanding of the challenges with regards to genre classification. The strategy used in this research integrates machine learning algorithms and windows forms applications as a user interface to make an impact on the music industry.

To ensure the results of this study are efficient, this study first looks at a Multilayer Perceptron (MLP) algorithm as a baseline model, which has an standard architecture of just 3 layers. The 3 layers comprise of an input, which receives pre-processed data, typically presented to the model as a feature vector, a hidden and output layer, where the model is trained to reduce errors between the target output and the computed values \cite{park2016artificial}. An example of the features used for a multilayer perceptron are Mel-Frequency Cepstral Coefficients which are mostly used for the purpose of music analysis.  The second algorithm is a CNN, which is a deep neural network that has multiple layers, with reduced parameters which provides researchers with the opportunity to work with larger models \cite{albawi2017understanding}. To name a few, the hyper-parameters typically used for these models are, namely batch sizes, epochs and learning rates. CNN models also make use of a split between the training, validation and test data. 

K-Nearest Neighbour is a classification model that classifies instances based on their similarities. It assigns a label to data point by identifying the class that is predominantly recognized within the nearest k-neighbours \cite{kataria2013review}. K-Nearest Neighbour models includes hyper-parameter tuning of the different values of k, to find the optimal value that provide the best classification result. The last model is a wider model, which is considered as a type of neural network that consists of nodes and layers. Wide models focus on the different patterns within the data. 

\begin{figure}
    \centering
    \includegraphics[width=.4\textwidth]{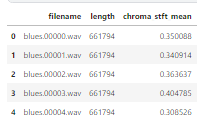}
    \caption{Metadata information retrieved from the dataset.}
    \label{fig:enter-label}
\end{figure}

The proposed dataset for this research is the GTZAN Dataset, which is applied specifically for the purpose of training machine learning models for classification. The dataset consists of 10 genres, with 100 audio files for each genre, that are 30 seconds long \cite{olteanu2020gtzan}. The different genres are "blues", "classical", "country", "disco", "hiphop", "jazz", "metal", "pop", "reggae", "rock". During the course of this research, I discovered that the dataset actually has 1 or 2 corrupted files that could not be processed, therefore entailing that those files had to be excluded from the dataset. The files in question were identified within the "jazz" genre.

\section{Data Pre-Processing}

Figure 3 is the Short-Time Fourier Transform (STFT) representation that is extracted from a randomly selected audio file. It contains the meaningful information that can be used to classify audio. It is basically the analysis of frequency data of a signal over short time intervals. Padding is applied to the extracted features, to ensure that the audio features that are extracted are of the same length, which is represented in figure 4. This also ensures that the model captures only the necessary information for the training process.

The process of STFT includes the division of the input audio signal into short frames, that overlap each other, capturing temporal information. Fourier Transform is then applied to each window frame, whereby a conversion from a time domain to a frequency domain occurs, which reveals the frequency of the signal in each frame. It is also key to highlight that STFT operates with fixed length segments (windows), therefore it may be necessary to apply padding, to ensure that the length of the signal remains constant during the transformation process. 

The GTZAN dataset also consists csv metadata that is utilized for the purpose of this research. The training data is split into 80\% for the training data, and 20\% for the testing phase. Figure 2 presents a sample of the metadata information, whereby a complete view of the metadata is a total of 60 columns which all contain information extracted from audio signals through STFT.

\begin{figure}
    \centering
    \includegraphics[width=.5\textwidth]{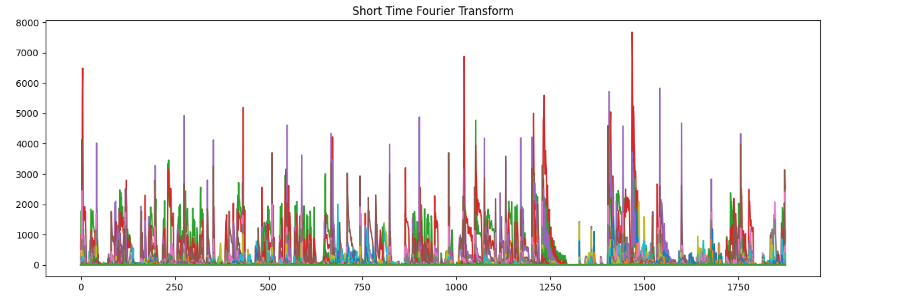}
    \caption{STFT of a randomly selected Reggae audio file}
    \label{fig:enter-label}
\end{figure}

\begin{figure}
    \centering
    \includegraphics[width=.5\textwidth]{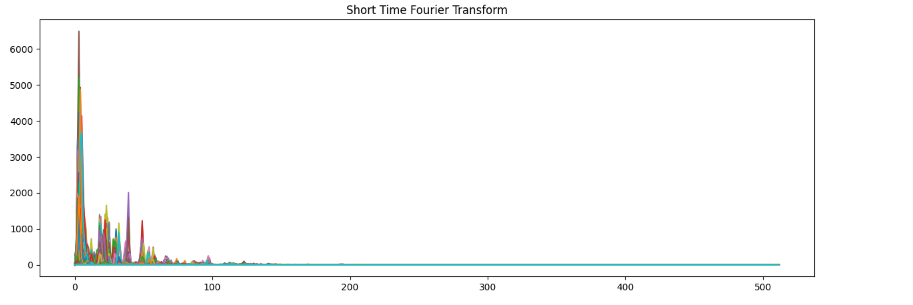}
    \caption{STFT of a randomly selected Reggae audio file in which padding has been applied}
    \label{fig:enter-label}
\end{figure}

The spectrogram representation which is obtained through STFT, contains the necessary features required for classification. The most important feature to highlight in this regard is the Mel-Frequency Cepstral Coefficients (MFCCs). MFCCS is typically used for speech recognition, speaker identification and musical genre classification. This feature contains the necessary frequency information that used to process audio signals of different genres which is important for classification. The numerical values determined from this features are converted into a 1D feature vector and provided as input into the models. Figure 5 represents the spectrogram.

\begin{figure}
    \centering
    \includegraphics[width=.5\textwidth]{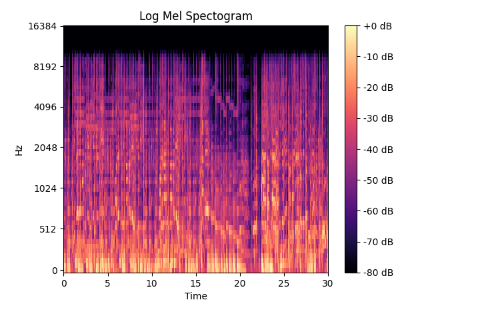}
    \caption{Spectrogram of the selected reggae genre audio file}
    \label{fig:enter-label}
\end{figure}

\begin{figure}
    \centering
    \includegraphics[width=.5\textwidth]{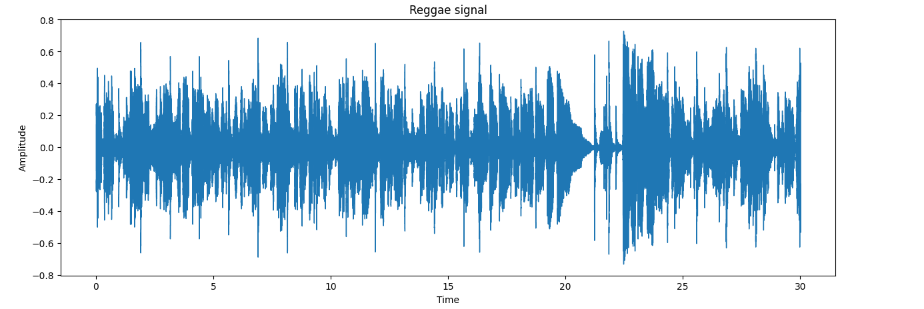}
    \caption{Wave form representation of a randomly selected Reggae audio file}
    \label{fig:enter-label}
\end{figure}

\section{Baseline Multilayer Perceptron}
\subsection{Model Specifications}
The MLP used in this research comprises of an input layer, a hidden layer with 1024 neurons, and an output layer in which a softmax function is applied to it. A ReLU activation function is applied, along with gradient descent with the purpose of fine-tuning hyperparameters.

In the context of this model, the complexity limits the model's capabilities to learn meaningful representations. The model has not been tested with techniques such as batch normalization, or drop out. The results presented by the model showcase that the prescribed idea proposed in this research of using an MLP with only 3 hidden layers for classification, is not an efficient solution.

The similarly related work by \cite{silla2008machine}, outlines the use of techniques such as feature selection, or a combination of computing space-time decomposition. In their research paper, the MLP's performance is able to produce results up to 56.40\%, which signifies the idea of using different techniques to improve performance \cite{silla2008machine}. 

\section{Deep Neural Network: CNN}
\subsection{Model Specifications and results}

The model consists of 7 layers. The first convolutional layer uses 32 filters with a kernel size of (3,3). The second convolutional layer consists of 64 filters, also with the kernel size of (3,3). A max pooling layer is included, and the output is then flattened to ensure that it is contained within a 1D array, which is the feature vector containing the necessary input values required for computation. The selected activation function is a Rectifier Linear Unit. When process the data, because the CNN uses the windows frames as input for performing classification, which are the time segments, the feature extraction process had to include padding, in which a specified max length is determined, and if the length of the time segment exceeds the max length, the values are truncated to fit the max length, or to add 0 values in a case where the time segments are not equal to the max length. This is done to ensure that the values remain constant for the CNN algorithm. 

In terms of calculating the loss, a categorical cross entropy function is used as it is most suitable for multiple output classifications. The selected optimizer is "adam" optimizer, with a softmax function being applied on the last layer of the model. The results from this architecture return a 0.25\% accuracy when the model is trained for 10 epochs, with a dense layer consisting of 128 neurons. The possible reason for this low accuracy rate could be that the model might benefit from an increase in complexity, which is what I further researched and noted, and indeed the accuracy rating increased.

When the model is provided with 2 dense layers of 256 and 128 neurons, along with a dropout rate of 0.3 which randomly sets the input values to 0 to prevent overfitting, its accuracy rate increases to around 0.275\%. This could mean that increasing the dense layers improves the performance of the model.

\begin{figure}
    \centering
    \includegraphics[width=.5\textwidth]{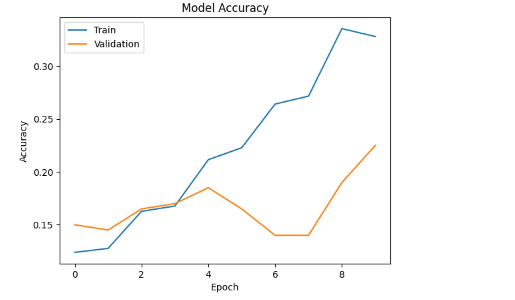}
    \caption{Training accuracy of CNN Model}
    \label{fig:enter-label}
\end{figure}

\begin{figure}
    \centering
    \includegraphics[width=.5\textwidth]{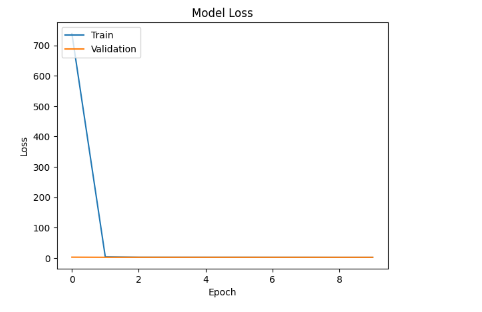}
    \caption{Training loss of CNN model.}
    \label{fig:enter-label}
\end{figure}

Figure 7 essentially represents an accuracy rating of about 23\%, in which the validation data shows that it is possible that, at some point, the model was not learning any information before a significant rise in accuracy. In comparison to the loss outlined in figure 8, this could mean the model was stuck in a local minima before continuing onwards. The change in accuracy is affected by the addition of dropout and an inclusion of another layer with 64 neurons. The model is trained for 10 epochs, whereby the increase in accuracy displayed on the chart, could signify that provided the training time is increased, the model accuracy could increase.

\section{K-Nearest Neighbours From Scratch}
\subsection{Model Specifications and results}
The K-Nearest model presented a decent result of 55\% for music genre classification. The models specifications include a K = 1 value, in which, increasing this value decreases the performance of the model significantly. 

K-fold cross validation is technique that can be used for improving the KNN's performance. A similar study highlighted the use of K-fold cross validation for performing classification \cite{tzanetakis2002musical}, which improves the performance, and also ensures that the accuracy is not biased. 

The KNN model is essentially a model that requires less training time compared to other existing models. Selecting a k value ranges from 1 to 100 in which the different values can be tested to identify the most optimal accuracy. The similarly related research work on genre classification presented a KNN with an accuracy rating of 77.18\%. The author opted for the Spotify dataset which contains all the values for the metadata of each genre \cite{rahardwika2020comparison}.

\section{Random Forest Wide model}
\subsection{Model Specifications and results}
The random forest is defined with 35 estimators and a depth of 25 for complexity. In this regard, the more complex the model is, the more it is able to interpret data and improve its accuracy, similarly for its number of estimators. The model's results are present a positive outcome of about 84\% for genre classification.

The confusion matrix entails that a lighter color essentially represent how well the model has learned meaningful representations. Interestingly, with regards to this particular model, the confusion matrix demonstrates the models ability to classify classical music most accurately compared to other genres. Figure 9 presents the confusion matrix.

\begin{figure}
    \centering
    \includegraphics[width=.5\textwidth]{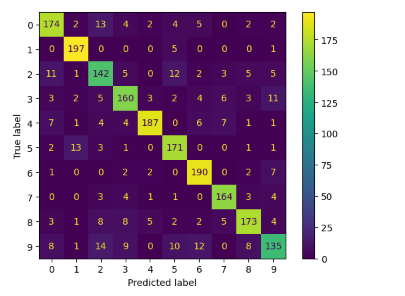}
    \caption{Confusion Matrix for Random Forest.}
    \label{fig:enter-label}
\end{figure}

\subsection{Comparisons}
\begin{tabular}{ | c | c | } 
  \hline
  \textbf{Model}  & \textbf{Results} \\
  \hline
  MLP & 0\% \\
  \hline
  CNN & 40\% \\
  \hline
  KNN & 55\% \\
  \hline
  RandomForest & 84\% \\
  \hline
  
\end{tabular}

\begin{figure}
    \centering
    \includegraphics[width=.2\textwidth]{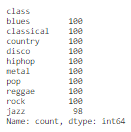}
    \caption{Count of the audio files for each genre}
    \label{fig:enter-label}
\end{figure}

\section{Conclusion}
To summarise the findings, the Random Forest model presented a higher accuracy rating compared to the MLP, CNN and KNN. The Random Forest's ensemble of decision trees allowed it to capture complex relationships in the data more effectively than the other models. In contrast, the MLP's architecture and limited capacity prevent the model from learning intricate patterns and useful information to perform classification. This is also similar to the covolutional neural network, whereby the models complexity plays a role in its ability to classify accurately. It is also possible that the CNN's fixed kernel sizes were not the most optimal solution for the purpose of this research.

In terms of feature engineering, STFT is a feature extraction method that can be applied to audio files to depict information that can be transposed into feature vectors, to train machine learning models. The dataset contains a few files that are corrupt, and it is possible that those files can impact the ability of the models to generalise when shown new data. The identified files were recognised in the "jazz" class, which could also result in a class imbalance compared to the rest of the classes. 

Figure 10 present a count of the number of audio files for each class that are processed and used for this research. The distribution of samples across classes provides considerations when interpreting the model's performance and generalization. To conclude this research, machine learning models present an avenue in music that can be further explored and utilized. 

\bibliographystyle{plain}
\bibliography{References}

\end{document}